\documentclass[journal]{IEEEtran}
\usepackage[latin1]{inputenc}
\usepackage{graphicx}
\usepackage{color}
\usepackage{placeins}
\usepackage{float}
\usepackage{tabularx,colortbl}
\usepackage{amssymb}
\usepackage{amsthm}
\usepackage{cite}
\usepackage{amsmath}
\usepackage{caption2}

\usepackage{cleveref}
\usepackage{multicol}       
\usepackage{multirow}       
\usepackage{array}          
\usepackage{makecell}
\usepackage{colortbl}
\usepackage{pifont}       
\usepackage{bbding}       
\usepackage{fontawesome}  
\usepackage{cases,subeqnarray}
\usepackage{bm,multirow,bigstrut}
\usepackage{textcomp}
\usepackage{latexsym,bm}
\usepackage{booktabs}
\usepackage{xcolor}
\usepackage{mathtools}
\usepackage{dsfont}
\usepackage{extarrows}
\usepackage{amsthm}

\usepackage{subfigure}
\usepackage{makecell}
\definecolor{lightblue}{rgb}{0.93,0.95,1.0}

\usepackage{algorithm} 
\usepackage{algorithmic}

\usepackage{booktabs} 
\usepackage{booktabs} 
\usepackage{longtable}
\usepackage{threeparttable}

\usepackage{custom_commands}

\theoremstyle{plain}

\theoremstyle{plain}

\theoremstyle{definition}

\IEEEoverridecommandlockouts

\begin{document}
\title{Channel Estimation for Rydberg Atomic Receivers}
\author{Bokai~Xu, Jiayi~Zhang,~\IEEEmembership{Senior Member,~IEEE}, Zhongtao~Chen, Bingyang Cheng,\\Ziheng~Liu, Yik-Chung~Wu,~\IEEEmembership{Senior Member,~IEEE}, and Bo~Ai,~\IEEEmembership{Fellow,~IEEE}

%

\thanks{B. Xu, J. Zhang, Z. Liu, and B. Ai are with School of Electronic and Information Engineering, Beijing Jiaotong University, Beijing, 100044, China (e-mails: 20251197@bjtu.edu.cn, zhangjiayi@bjtu.edu.cn, boai@bjtu.edu.cn); Z. Chen, B. Cheng, and Y. Wu are with the Department of Electrical and Electronic Engineering, The University of Hong Kong, Hong Kong (e-mails: ztchen@eee.hku.hk, bycheng@eee.hku.hk, ycwu@eee.hku.hk).}
}
\maketitle


\begin{abstract}
The rapid development of the quantum technology presents huge opportunities for 6G communications. Leveraging the quantum properties of highly excited Rydberg atoms, Rydberg atom-based antennas present distinct advantages, such as high sensitivity, broad frequency range, and compact size, over traditional antennas. To realize efficient precoding, accurate channel state information is essential. 
However, due to the distinct characteristics of atomic receivers, traditional channel estimation algorithms developed for conventional receivers are no longer applicable. To this end, we propose a novel channel estimation algorithm based on projection gradient descent (PGD), which is applicable to both one-dimensional (1D) and two-dimensional (2D) arrays.
Simulation results are provided to show the effectiveness of our proposed channel estimation method.
\end{abstract}

\IEEEpeerreviewmaketitle
\section{Introduction}
The relentless demand for higher data rates, enhanced user experiences, and ubiquitous connectivity has propelled the evolution from fifth-generation (5G) to the forthcoming 6G wireless communication networks \cite{10556753, xu2023resource}. In recent years, there has been a growing interest in leveraging atomic-scale technologies in multiple-input-multiple-output (MIMO) communications. Termed as atomic MIMO receivers \cite{cuifrontier, gong2024}, these innovative systems utilize atomic-level interactions to achieve unprecedented levels of efficiency and performance in wireless communications \cite{QuanComm_Wang2022}.

Atomic MIMO receivers integrate advanced materials and nanostructures with quantum-based principles, enabling the manipulation of electromagnetic (EM) waves at the atomic scale. By employing techniques such as coherent detection and adaptive beamforming, atomic MIMO systems can dynamically adjust to varying channel conditions \cite{gong2024, liu_deep_2022, cuifrontier, QuanComm_Wang2022}.
Rydberg atoms receivers, which rely on electron transitions between different energy levels when encountering an EM wave \cite{AtomicTrans}, demonstrate superior detection and sensing accuracies, achieving sub-millimeter-level granularity, while traditional receivers operate at a coarser granularity. Moreover, unlike traditional antennas that require costly arrays to match various frequency bands, atomic MIMO receivers can efficiently respond to a broad frequency range with fewer resources, thus reducing overall system complexity and cost \cite{liu_deep_2022, rehman2024}.

Due to the unique benefit of atomic MIMO receivers, theoretical modeling and signal processing for atomic receivers 
have appeared recently. For example, 
\cite{kim2024} proposed a multiple signal classification (MUSIC) algorithm for quantum wireless sensing of multiple users. Moreover, a biased Gerchberg-Saxton (GS) algorithm was proposed to address the non-linear biased phase retrieval (PR) problem caused by the magnitude-only measurements of atomic receivers \cite{cuireceivers}. In  \cite{cui2024}, a MIMO precoding design for atomic receivers that utilizes independent processing of In-phase-and-Quadrature (IQ) data, was introduced. Unfortunately, up to now, there are no related works on channel estimation for atomic MIMO receivers, which is essential for effective signal detection and precoding. Furthermore, compared to the linear model used in conventional MIMO systems, the nonlinear transmission model significantly complicates channel estimation.

\begin{figure}[t]
	\centering
	\includegraphics[width=0.45\textwidth]{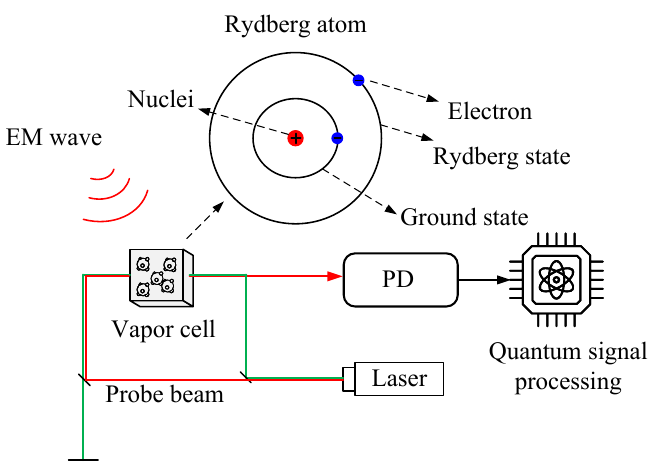}
	\caption{Schematic representation of the atomic receiver using Rydberg atoms. The system encompasses a vapor cell that interacts with an EM wave to excite Rydberg states. The resulting RF signals are detected and processed via a photo-detector (PD) and a quantum signal processing unit.}
    \label{Schematic_atom}
\end{figure}

To fill this gap, we for the first time propose a channel estimation framework for Rydberg atomic receivers, applicable to both 1D and 2D antenna arrays. Specifically, we reformulate the channel estimation problem as a matrix rank constrained optimization problem, and propose a novel channel estimation algorithm based on the projection gradient descent (PGD). Compared with the conventional GS algorithm for addressing phase recovery challenges, our proposed algorithm not only delivers comparable performance in 1D case, but also applicable to 2D antenna arrays. Simulation results show the superior performance of the proposed algorithm.

\section{Preliminaries of Rydberg Atomic Receivers}

\begin{figure}[t]
	\centering
	\includegraphics[width=0.45\textwidth]{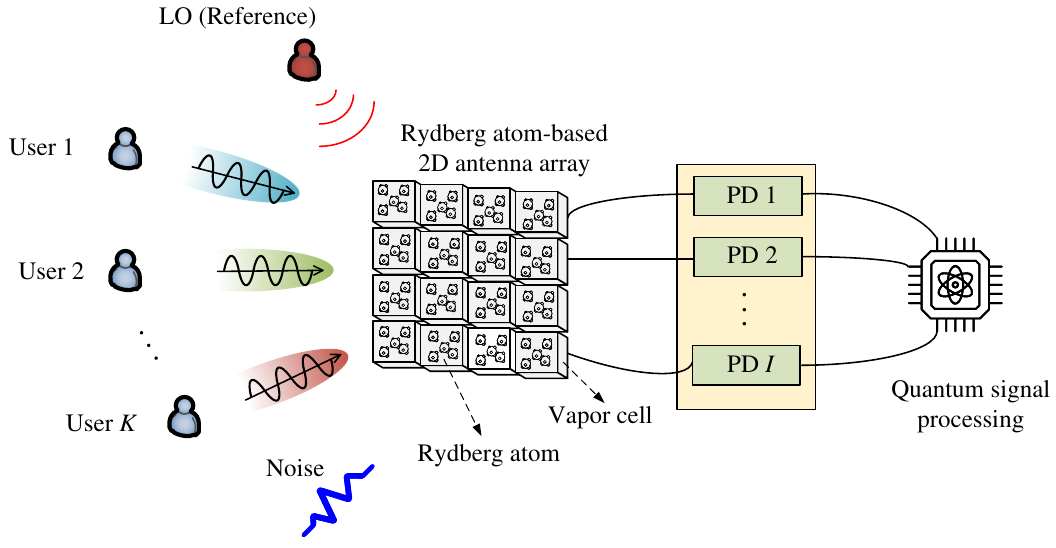}
	\caption{An illustration of the Rydberg atom-based 2D antenna array system with $K$ users and a reference.}
    \label{System_Model}
\end{figure}

As shown in Fig. \ref{System_Model}, we consider a MIMO wireless communication system with $K$ single-antenna users, and the base station (BS) is equipped with an atomic array and receiver. 
We first consider a Rydberg atom-based 1D antenna array at the BS, which comprises $I$ vapor cells, and each is connected to a photodetector (PD). These vapor cells are uniformly distributed along a linear dimension at the receiver side. 
Rydberg atoms exhibit unique quantum behaviors when transitioning from the ground state to excited states due to absorption of EM waves. This transition can be mathematically characterized using a two-level quantum system representation. 

The general state of a Rydberg atom in the $i$-th vapor cell at time $t$ can be expressed as
\begin{equation}
| \Phi(t) \rangle_i = \zeta_{g,i}(t) |g\rangle + \zeta_{e,i}(t) |e\rangle,
\end{equation}
where $\zeta_{g,i}(t)$ and $\zeta_{e,i}(t)$ are the amplitudes for the ground and excited states at time $t$, respectively, constrained by the condition $|\zeta_{g,i}(t)|^2 + |\zeta_{e,i}(t)|^2 = 1$.
The ground state $|g\rangle$ and the excited state $|e\rangle$ are defined as $|g\rangle=[0,1]^{T}$ and $|e\rangle=[1,0]^{T}$, respectively.
In general, the state equation 
is governed by the Schr\"{o}dinger equation \cite{QuantumMechanism_Zwiebach2006}
\begin{equation}
i \hbar \frac{\partial | \Phi(t) \rangle_i}{\partial t} = ( \hat{H}_i + \hat{V}_i ) | \Phi(t) \rangle_i,
\end{equation}
where $\hbar=h/2 \pi$ is the reduced Planck constant and $h=6.626 \times 10^{-34}$ J$\cdot$s is the Planck constant. \( \hat{H}_i \) denotes the free Hamiltonian defined as $\hat{H}_i = \text{diag}(\hbar \omega_e, \hbar \omega_g)$, where
\( \omega_e \) and \( \omega_g \) represent the angular frequencies of the excited and ground states, respectively. The interaction Hamiltonian \( \hat{V}_i \) encapsulates the energy dynamics of the EM wave interacting with the atom. In particular, 
$\hat{V}_i = \varpi_i(t) |e\rangle \langle g| + \varpi_i^*(t) |g\rangle \langle e|$,
with
\begin{equation}
\label{varpi}
\varpi_i(t) = \sum_{k=1}^{K}\sum_{l=1}^{L_{k}} \boldsymbol{\mu}_{eg}^{T} \boldsymbol{\epsilon}_{i,k,l}\alpha_{l,k} s_k \cos\left(\omega t + (i-1) \varphi_{l,k}\right),
\end{equation}
where $L_{k}$ is the number of paths for the $k$-th user, $\alpha_{l,k}$ is the path gain of the $l$-th path for the $k$-th user, and $s_k$ is the information bearing symbol or pilot symbol from the $k$-th user.
The phase shift for a radio wave propagating from the transmit antenna to the $i$-th atomic antenna via the $l$-th path is given by $\varphi_{l,k} = \frac{2 \pi d}{\lambda} \cos \theta_{l,k}$, where $d$ is the inter-antenna spacing, $\lambda$ is the wavelength, and $\theta_{l,k}$ is the angle-of-arrival (AOA).
In equation \eqref{varpi}, $\boldsymbol{\mu}_{eg}$ is the transition dipole moment, and $\boldsymbol{\epsilon}_{i,k,l}$ represents the polarization direction.
The frequency $\omega$ of the electromagnetic wave is tuned to be near the transition frequency $\omega_{eg} = \omega_e - \omega_g$, ensuring that quantum jumps predominantly occur between these two energy levels.
Note that the EM wave at $i$-th vapor cell
\begin{equation}
\mathbf{E}_{i}(t)=\sum_{k=1}^{K}\sum_{l=1}^{L_{k}} \boldsymbol{\epsilon}_{i,k,l} \alpha_{l,k} s_{k}\cos\left(\omega t + (i-1) \varphi_{l,k}\right)
\end{equation}
is encoded in $\varpi_i(t)$.

The amplitude of the excited state \( |\zeta_{e,i}(t)|^2 \) is given by the Rabi frequency at the \( i \)-th vapor cell \( \Omega_i \) \cite{gong2024}
\begin{equation}
|\zeta_{e,i}(t)|^2 = \sin^2\left(\frac{\Omega_i}{2} t\right),
\end{equation}
where
\begin{equation}
\Omega_i = \frac{1}{\hbar} \left| \sum_{k=1}^{K}\sum_{l=1}^{L_{k}} \left( \boldsymbol{\mu}_{e g}^{H} \boldsymbol{\epsilon}_{i, k,l} \right) \alpha_{l,k} s_k e^{j (i-1)\varphi_{l,k}} \right|.
\end{equation}
The function of the atomic-MIMO receiver is to extract the values of the Rabi frequency $\Omega_{i}$ and 
we denote the measurement of $\Omega_i$ as the received signal at the $i$-th antenna.

Based on the analysis in \cite{9748947, cuireceivers}, the channel between the $i$-th vapor cell and the $k$-th user is modeled as
\begin{align}
\label{channel_1D}
    g_{i,k} = \sum_{l=1}^{L_{k}} \frac{1}{\hbar} \boldsymbol{\mu}_{eg}^{T} \boldsymbol{\epsilon}_{i,k,l} \alpha_{l,k} e^{-j (i-1) \varphi_{l,k}}.
\end{align}
Given transmitted signal $s_{k,p}$ from the $k$-th user at the $p$-the time slot, the received signal is \cite{9748947, cuireceivers}
\begin{align}
    y_{i,p} = \left| \sum_{k=1}^{K} g_{i,k} s_{k,p} + b_{i,p} + n_{i,p} \right|, \label{eq:y}
\end{align}
where $\left| \cdot \right|$ denotes the magnitude of the argument, and
\begin{align}
b_{i,p}=\frac{s_{b,p}}{\hbar}\boldsymbol{\mu}_{eg}^{T}\boldsymbol{\epsilon}_{b,i} \alpha_{b} e^{-j (i-1) \varphi_{b}}
\end{align}
is a known reference signal composed of the $s_{b,p}$ passes through a channel with path gain $\alpha_b$, phase shift $\varphi_b$ and polarization $\boldsymbol{\epsilon}_{b,i}$, while the noise $n_{i,p}$ follows Gaussian distribution $\mathcal{C N}\left(0, \sigma^{2} \right)$. Estimating $g_{i,k}$ using \eqref{eq:y} can be viewed as a phase retrieval problem and the Gerchberg Saxton (GS) algorithm \cite{7130654} can be used. However, when the antenna array is extended to multiple dimensions (to be detailed in the next section), the aforementioned method becomes inapplicable.

To develop a unified channel estimation framework, we consider the case where the reference-to-receiver distance is much shorter than the transmitter-to-receiver distance\footnote{In practice, a reference signal generated by a nearby, controlled source (like a local transmitter or dedicated antenna) travels a short distance to the receiver. This ensures its amplitude dominates, as other signals typically experience greater path loss and fading over longer distances.}, which leads to $\left | b_{i,p} \right |\gg \left | \sum_{k=1}^{K} g_{i,k} s_{k,p} + n_{i,p} \right |$.
Defining $a_{i,p}=\sum_{k=1}^{K} g_{i,k} s_{k,p}$, the received signal is rewritten as
\begin{equation}
\label{appro_proof}
\begin{aligned}
y_{i,p}&=\left | a_{i,p}+b_{i,p}+n_{i,p} \right |\\
&=\left |b_{i,p}\right |\sqrt{1+\frac{\left |a_{i,p}+n_{i,p}\right |^2}{\left | b_{i,p} \right |^2 }+2\operatorname{Re}\left(\frac{a_{i,p}+n_{i,p}}{b_{i,p}} \right) }\\ 
&\stackrel{(a)}{\approx}\left |b_{i,p}\right |\left(1+\frac{\left |a_{i,p}+n_{i,p}\right |^2}{2\left | b_{i,p} \right |^2 }+\operatorname{Re}\left(\frac{a_{i,p}+n_{i,p}}{b_{i,p}} \right)\right),
\end{aligned}
\end{equation}
where $(a)$ comes from the Taylor expansion $\sqrt{(1+x)}=1+\frac{1}{2} x+\mathcal{O}\left(x^{2}\right)$ under $\left | b_{i,p} \right |\gg \left | a_{i,p}+n_{i,p}\right |$.
Thus, \eqref{eq:y} is simplified to
\begin{align}
\label{y_{i,p}}
    y_{i,p} = \operatorname{Re}\left( e^{-j z_{i,p}}  \sum_{k=1}^{K} g_{i,k} s_{k,p} \right) + \left| b_{i,p} \right| + \bar{n}_{i,p},
\end{align}
where $z_{i,p}$ is the angle of $b_{i,p}$, and $\bar{n}_{i,p}=\operatorname{Re}\left(e^{-j\angle b_{i,p}}n_{i,p}\right)$ follows Gaussian distribution $\mathcal{N}\left(0, \sigma^{2}/2 \right)$.


Arranging \eqref{y_{i,p}} with different $i$ and $p$ into a matrix, we obtain
\begin{align}
    \m{Y} = \operatorname{Re}\left( \m{Z} \circ \left( \m{G} \m{S} \right) \right) + \left| \m{B} \right| +\m{N}, \label{eq:y_approx}
\end{align}
where $\m{Y} \in \R^{I \times P}$, $\m{Z} \in \C^{I \times P}$, $\m{B} \in \C^{I \times P} $ and $\m{N} \in \C^{I \times P}$ are with the $(i,p)$-th element being $y_{i,p}$, $e^{-j z_{i,p}}$, $b_{i,p}$ and $\bar{n}_{i,p}$, respectively. Furthermore, $\mathbf{G} \in \C^{I \times K}$ with its $(i,k)$-th element being $g_{i,k}$ and $\m{S} \in \C^{K \times P}$ with its $(k,p)$-th element being $s_{k,p}$. Furthermore, 
$\circ$ denotes Hadamard product, and $\left | \mathbf{B} \right | $ denotes elementwise absolute value of $\mathbf{B}$.

\section{Channel Estimation from 1D to 2D Antenna Array}
To realize efficient channel estimation in Rydberg atomic receivers with a unified framework for 1D and 2D arrays, we propose a gradient based method in this section.

\subsection{Channel Estimation for 1D Antenna Array}
Mathematically, the channel estimation problem based on \eqref{eq:y_approx} can be formulated as solving the following problem\footnote{We assume perfect synchronization and prior coordination between the transmitter and receiver for pilot transmission.}
\begin{equation}
\argmin _{\mathbf{G}}\mathcal{L}(\mathbf{G})=\|\mathbf{Y}-|\mathbf{B}|-\operatorname{Re}(\mathbf{Z} \circ(\mathbf{G S}))\|_{F}^2.
\end{equation}
Then, the objective function is minimized via gradient descent, i.e.,
\begin{equation}
\mathbf{G}^{t+1}=\mathbf{G}^{t}-\eta^{t} \nabla_{\mathbf{G}^{*}} \mathcal{L}\left(\mathbf{G}^{t}\right),
\end{equation}
where $\eta^{t}$ is the step-size, and 
\begin{equation}
\nabla_{\mathbf{G}} \mathcal{L}\left(\mathbf{G}^{t}\right)=-2\left(\left(\mathbf{Y}-\left |\mathbf{B}\right |-\operatorname{Re}\left(\left(\mathbf{G}^{t}\mathbf{S}\right) \circ  \mathbf{Z}\right)\right) \circ \mathbf{Z}^{\ast }\right) \mathbf{S}^{H}.
\end{equation}
While the above solution seems straightforward, it becomes more meaningful if the antenna array is 2D. 
\subsection{Extension to 2D Antenna Array}
In this case, the atomic MIMO receiver is equipped with vapor cells arranged in a 2D planar array, with size $I_1 \times I_2$. Consequently, the channel model is extended from \eqref{channel_1D} and becomes 
\begin{align}
\label{gik}
g_{i_1,i_2,k} = \sum_{l=1}^{L_{k}} \frac{1}{\hbar}\boldsymbol{\mu}_{eg}^{T}\boldsymbol{\epsilon}_{i_1,i_2,k,l} \alpha_{l,k} e^{-j\left[(i_1-1) u_{l,k}+(i_2-1) v_{l,k}\right]},
\end{align}
where $i_{1}$, $i_{2}$ are the antenna indices, $u_{l,k}=\frac{2 \pi d_{1}}{\lambda} \cos \theta_{l,k}$, $v_{l,k}=\frac{2 \pi d_{2}}{\lambda} \sin \theta_{l,k} \cos \phi_{l,k}$ represent the phase shift with $\theta_{l,k}$ and $\phi_{l,k}$ representing the elevation angle and azimuth angle of the $l$-th path of the $k$-th user, respectively.
We define the received reference signal $b_{i_1,i_2,p}$ as
\begin{align}
b_{i_1,i_2,p}=\frac{s_{b,p}}{\hbar}\boldsymbol{\mu}_{eg}^{T}\boldsymbol{\epsilon}_{b,i_1,i_2} \alpha_{b} e^{-j\left[(i_1-1) u_{b}+(i_2-1) v_{b}\right]}.
\end{align}
Similar to \eqref{y_{i,p}}, the received signal can be derived to be 
\begin{align}
\label{y_2D}
    y_{i_1,i_2,p} = \operatorname{Re}\left( e^{-j z_{i_1,i_2,p}}  \sum_{k=1}^{K} g_{i_1,i_2,k} s_{k,p} \right) + \left| b_{i_1,i_2,p} \right| + n_{i_1,i_2,p},
\end{align}
which can be put in a $I_1 \times I_2 \times P$ tensor as 
\begin{align}
    \ten{Y} = \operatorname{Re}\left( \ten{Z} \circ \left( \ten{G} \times_3 \m{S}^{T} \right) \right) + \left| \ten{B} \right| + \ten{N},
\end{align}
where $\ten{G} \times_3 \m{S}^{T}$ is the multiplication of the matrix $\m{S}^{T}$ to the third mode of the tensor $\ten{G}$.
Notice that for the three dimensional tensor $\ten{G}$, each frontal slice represents the channel matrix of a particular user $k$. According to \eqref{gik}, since there are only $L_k$ paths for user $k$'s channel, the $k$-th frontal slice of $\ten{G}$ has only rank $L_k$. This is in contrast to 1D case where the channel of each user is represented by a vector, and such rank property does not exist.

To estimate the channel, we apply mode-3 unfolding to $\ten{Y}$:
\begin{align}
\label{tensor}
    \m{Y}_{(3)} = \operatorname{Re}\left( \m{Z}_{(3)} \circ ( \m{S}^{T} \m{G}_{(3)} ) \right) + \left| \m{B}_{(3)} \right| + \m{N}_{(3)}.
\end{align}
Therefore, the optimization problem for channel estimation is 
\begin{align}
\label{optimization_2D}
    &\min_{\m{G}_{(3)}} ~~ \norm{\m{Y}_{(3)} - \left| \m{B}_{(3)} \right| - \operatorname{Re}\left( \m{Z}_{(3)} \circ ( \m{S}^{T} \m{G}_{(3)} ) \right)}_F^2,  \\
    &~~\mathrm{s.t.} ~~ \mathrm{rank} \left( \mathrm{mat} \left( [ \m{G}_{(3)} ]_{k,:} \right) \right) \le  L_{k}, \forall k=1,2, \ldots, K, \nonumber
\end{align}
where $\operatorname{mat}([\m{G}_{(3)}]_{k,:})$ means constructing a matrix from the $k$-th row of $\m{G}_{(3)}$,  which corresponds to the $k$-th frontal slice of $\ten{G}$ with size $I_1 \times I_2$.
Note that compared to 1D array, the channel recovery in 2D array includes a rank constraint of $\mathbf{G}_{(3)}$.  
This makes conventional phase retrieval algorithms not applicable here. 
Moreover, modifying GS to account for the rank constraint is
non-trivial because the iterative projections in GS are tailored
to satisfy magnitude constraints in the signal domain and do
not naturally extend to enforcing low-rank properties in the
channel domain.

We define the objective function in \eqref{optimization_2D} as
\begin{align}
\label{f}
f(\mathbf{G}_{(3)})\!=\!\norm{\m{Y}_{(3)} \!-\! \left|\! \m{B}_{(3)} \!\right| \!-\! \operatorname{Re}\left(\! \m{Z}_{(3)}\! \circ \!\left(\! \m{S}^{T} \m{G}_{(3)}\! \right)\! \right)}_F^2.
\end{align}
To minimize \eqref{f} under the rank constraint, we propose to use the projected gradient descent (PGD) method:
\begin{equation}
\label{G_3}
\mathbf{G}_{(3)}^{t+1}= \operatorname{proj}_{\mathcal{X}}\left(\mathbf{G}_{(3)}^{t}-\zeta ^{t} \nabla_{\mathbf{G}_{(3)}} f\left(\mathbf{G}_{(3)}^{t}\right)\right),
\end{equation}
where 
\begin{equation}
\begin{aligned}
\label{gradient}
&\nabla_{\mathbf{G}_{(3)}} f\left(\mathbf{G}_{(3)}^{t}\right)\\&=-\mathbf{S}^{\ast }\left(\left(\mathbf{Y}_{(3)}\!-\!\left |\mathbf{B}_{(3)}\right |\!-\!\operatorname{Re}\left( \m{Z}_{(3)}\! \circ\! \left( \m{S}^{T} \m{G}_{(3)}\! \right) \right)\right) \circ \mathbf{Z}_{(3)}^{\ast }\right),
\end{aligned}
\end{equation}
$\mathcal{X}=\{\mathbf{X} \in \mathbb{R}^{K \times I_{1} I_{2}} \mid \operatorname{rank}(\operatorname{mat}([ \mathbf{X} ]_{k,:})) \leq L_{k}, \forall k\}$ and $\zeta ^{t}$ is the step size. 

In \eqref{G_3}, it involves a projection operation. In general, $ \operatorname{proj}_{\mathcal{X}}(\mathbf{M})$ means solving the problem:
\begin{equation}
\label{projection}
\begin{aligned}
\underset{\mathbf{X}}{\operatorname{min}}\|\mathbf{X}-\mathbf{M}\|_{F}^{2} \\
\text { s.t. } \mathbf{X} \in \mathcal{X}.
\end{aligned}
\end{equation}
Performing singular value decomposition (SVD) on $\operatorname{mat}(\left[ \mathbf{M}\right]_{k,:})$ yields \cite{8918708}
\begin{equation}
\label{U_k}
\mathbf{U}_{k} \mathbf{\Sigma}_{k} \mathbf{V}_{k}^{T}=\operatorname{svd}(\operatorname{mat}(\left[ \mathbf{M}\right]_{k,:})).
\end{equation}
According to the definition of $\mathcal{X}$, the projection for each row of $\mathbf{M}$ is separable. Therefore, using Eckart-Young-Mirsky theorem, the solution to \eqref{projection} is
\begin{equation}
\label{X_k}
\left[ \mathbf{X}\right]_{k,:}=\operatorname{vec}\left(\widetilde{\mathbf{U}}_{k} \widetilde{\mathbf{\Sigma}}_{k} \widetilde{\mathbf{V}}_{k}^{T}\right)^T,
\end{equation}
where $\widetilde{\mathbf{U}}_{k}=\left[\mathbf{U}_{k}\right]_{:,1: L}$,
$\widetilde{\mathbf{V}}_{k}=\left[\mathbf{V}_{k}\right]_{:,1: L}$, and $\widetilde{\mathbf{\Sigma }}_{k}=\left[\mathbf{\Sigma}_{k}\right]_{1: L,1: L}$. The PGD iterative channel estimation algorithm is summarized in Algorithm~\ref{alg:1},
with the complexity being $\mathcal{O}(T[K \min \left(I_{1} I_{2}^{2}, I_{1}^{2} I_{2}\right)+P K I_{1} I_{2}])$, where $T$ is the number of iterations.
For 1D arrays, the complexity reduces to $\mathcal{O}(T P K I)$, since only gradient descent is applied and the projected step is not needed.
Under the assumption of Lipschitz continuous gradients, the algorithm guarantees a sufficient decrease condition for the objective function and convergence to a stationary point within $O\left(1 / \epsilon\right)$ iterations for a given accuracy $\epsilon$ \cite{8918708}.

\begin{algorithm}[!t] 
	\caption{The Proposed PGD Algorithm} 
	\label{alg:1} 
	\begin{algorithmic}[1] 
       \REQUIRE Mode-3 unfolded tensor $\mathbf{Y}_{(3)}$, $\mathbf{Z}_{(3)}$ and $\mathbf{B}_{(3)}$; rank $L$; $\mathbf{S}$
       \STATE Initialize the system parameters and optimization variables.
       \REPEAT
       \STATE Update gradient $\nabla_{\mathbf{G}_{(3)}} f\left(\mathbf{G}_{(3)}^{t}\right)$ based on \eqref{gradient}.
       \STATE Update $\mathbf{M}=\mathbf{G}_{(3)}^{t}-\zeta ^{t} \nabla_{\mathbf{G}_{(3)}} f\left(\mathbf{G}_{(3)}^{t}\right)$.
       \FOR{$k=1, \ldots, K$}
       \STATE Update $\left[ \mathbf{X}\right]_{k,:}$ using \eqref{X_k}.
       \ENDFOR
       \STATE Set $\mathbf{G}_{(3)}^{t+1}=\mathbf{X}$, $t=t+1$.
      \UNTIL $||\mathbf{G}_{(3)}^{t+1} - \mathbf{G}_{(3)}^{t} ||_F^2$ becomes smaller than a pre-defined threshold.
     \RETURN Channel matrix $\mathbf{G}_{(3)}$
      
	\end{algorithmic}
\end{algorithm}
\section{Cram\'{e}r-Rao Lower Bound Analysis}
We now derive the Cram\'{e}r-Rao Lower Bound (CRLB) to obtain a benchmark for the accuracy of channel estimation. 
First, we rewrite \eqref{tensor} as
\begin{equation}
\label{y_vec}
\overline{\mathbf{y}}=\operatorname{Re}\left(\overline{\mathbf{z}}\circ (\mathbf{I}\otimes \mathbf{S}^{T})\overline{\mathbf{g}} \right)+\left|\overline{\mathbf{b}}\right|+\overline{\mathbf{n}},
\end{equation}
where $\overline{\mathbf{y}}=\operatorname{vec}(\boldsymbol{\mathcal{Y}}) \in \mathbb{R}^{I_{1}I_{2}P\times 1}$,
$\overline{\mathbf{z}}=\operatorname{vec}(\boldsymbol{\mathcal{Z}}) \in \mathbb{C}^{I_{1}I_{2}P\times 1}$,
$\overline{\mathbf{g}}=\operatorname{vec}(\boldsymbol{\mathcal{G}}) \in \mathbb{C}^{I_{1}I_{2}K\times 1}$,
$\overline{\mathbf{b}}=\operatorname{vec}(\boldsymbol{\mathcal{B}}) \in \mathbb{C}^{I_{1}I_{2}P\times 1}$,
$\overline{\mathbf{n}}=\operatorname{vec}(\boldsymbol{\mathcal{N}}) \in \mathbb{R}^{I_{1}I_{2}P\times 1}\sim \mathcal{N}\left(\mathbf{0}, \sigma^{2} \mathbf{I}_{I_{1}I_{2}P}\right)$, 
and $\otimes$ denotes the Kronecker product.
Then, the conditional probability density function
 of $\overline{\mathbf{y}}$ with the given $\overline{\mathbf{g}}$ is

\begin{equation}
\begin{aligned}
&\Pr\left({\overline{ \mathbf{y}} \mid \overline{\mathbf{g}}}\right)\\&=\frac{1}{\left(2 \pi \sigma^{2}\!\right)^{I_{1}\!I_{2}\!P}} \exp\! \left\{-\frac{1}{2 \sigma^{2}}\!\left\|\overline{\mathbf{y}}-\!\left|\overline{\mathbf{b}}\right|\!-\operatorname{Re}\!\left(\overline{\mathbf{z}}\circ (\mathbf{I}\!\otimes\! \mathbf{S}^{T})\overline{\mathbf{g}} \right)\right\|_{2}^{2}\right\}.
\end{aligned}
\end{equation}
The Fisher information matrix (FIM) of \eqref{y_vec} can then be derived as
\begin{equation}
\begin{aligned}
\mathbf{FIM}(\overline{\mathbf{g}} )&=-\mathrm{E}\left\{\frac{\partial^{2}\ln(\Pr({\overline{ \mathbf{y}}) \mid \overline{\mathbf{g}}})}{\partial \overline{\mathbf{g}}^{\ast } \partial \overline{\mathbf{g}}}\right\}\\
&=\frac{1}{4 \sigma^{4}} (\mathbf{I} \otimes \mathbf{S}^{T})^{H} (\sigma^{2}\mathbf{I}\circ(\overline{\mathbf{z}}^{*}\overline{\mathbf{z}}^{T}))(\mathbf{I} \otimes \mathbf{S}^{T}).
\end{aligned}
\end{equation}
Since each element in $\overline{\mathbf{z}}$ takes the form $e^{-j \angle b_{i, p}}$, the matrix $\mathbf{I} \circ\left(\overline{\mathbf{z}}^{*} \overline{\mathbf{z}}^{T}\right)=\mathbf{I}$, the CRB becomes 
\begin{equation}
\label{C_R_B}
\begin{aligned}
\mathbf{C R B}\left(\overline{\mathbf{g}}\right)&\succeq \left(\mathbf{FIM}(\overline{\mathbf{g}} )\right)^{-1}\\
&=4 \sigma^{2}\left(\mathbf{I}_{I_{1} I_{2}} \otimes\left(\mathbf{S}^{*}\mathbf{S}^{T} \right)^{-1}\right).
\end{aligned}
\end{equation}
From \eqref{C_R_B}, it can be observed that the CRLB does not change with the number of antennas, decreases with signal-to-noise ratio (SNR), and reduces as the number of pilots grows.

\begin{figure}[t]
	\centering
	\includegraphics[width=0.33\textwidth]{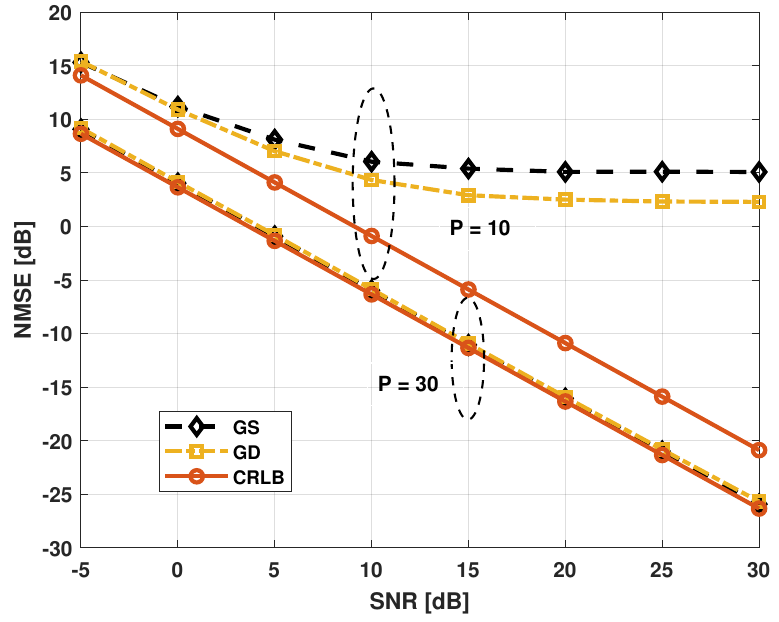}
	\caption{NMSE performance comparison with respect to the SNR under Rydberg atom-based 1D antenna array.}
    \label{NMSE_SNR_1D}
\end{figure}

\section{Simulation Results}
In the simulation, the Rydberg energy levels \(52D_{5/2}\) and \(53P_{3/2}\) (highly excited quantum states) are utilized to detect RF signals corresponding to the angular transition frequency \(\omega_{eg} \approx 2\pi \times 5 \, \text{GHz}\). Using the ARC package \cite{SIBALIC2017319}, the transition dipole moment $\boldsymbol{\mu}_{eg}$ for these levels is represented as $\boldsymbol{\mu}_{eg} = [0, 1785.9 \, q a_0, 0]^T$,
where \(q = 1.602 \times 10^{-19} \, \text{C}\) is the charge of an electron, and \(a_0 = 5.292 \times 10^{-11} \, \text{m}\) is the Bohr radius. The polarization direction is randomized, with $\boldsymbol{\epsilon}_{i,k,l}$, $\boldsymbol{\epsilon}_{b,i}$ following \(\mathcal{N}(0, \frac{1}{3})\). The channel gain $\alpha_{l,k}$ and $\alpha_{b}$ follow $\mathcal{CN}(0, 1)$ and $\mathcal{CN}(0, 10)$, respectively.
To assess the accuracy of various methods, the normalized mean square error (NMSE) performance, defined as
$\text{NMSE} = \frac{\mathrm{E}\{\| \mathbf{G}_{(3)} - \hat{\mathbf{G}}_{(3)} \|_F^2\}}{\mathrm{E}\{\| \mathbf{G}_{(3)} \|_F^2\}}$,
where $\hat{\mathbf{G}}_{(3)}$ is an estimate of the channel, is employed.
We set the number of users to $K=3$. 
It is assumed that each user has the random number of channel paths, i.e., $L_{k} \sim \operatorname{Uniform}(3,7)$.
The pilot symbols are independent and identically distributed and are generated from $\mathcal{C N}(0, 1)$.  
The PGD algorithm initializes the elements of channel matrix $\mathbf{G}_{(3)}^{0} \in \mathbb{C}^{K \times I_{1} I_{2}}$ as i.i.d. random complex-valued variables, i.e., $[\mathbf{G}_{(3)}^{0}]_{k, i} \sim \mathcal{C N}(0, 0.1)$.


The first simulation presents the NMSE performance under different SNRs for a Rydberg atom-based 1D antenna array with $I=8$ vapor cell elements.
We can see that when the size of the pilot is relatively small (e.g., $P=10$), although the proposed GD method performs better the GS algorithm \cite{cuireceivers, 7130654}, they perform far from the CRLB. This is because, while the GS algorithm does not rely on an approximation of the channel transmission model, it fails to account for the impact of noise. On the other hand, in cases with a high number of pilots (e.g., $P=30$), both the proposed GD and the GS algorithm perform almost identically and close to the CRLB. This shows the proposed GD algorithm could be a good contender to replace GS algorithm in Rydberg receiver.

\begin{figure}[t]
	\centering
	\includegraphics[width=0.33\textwidth]{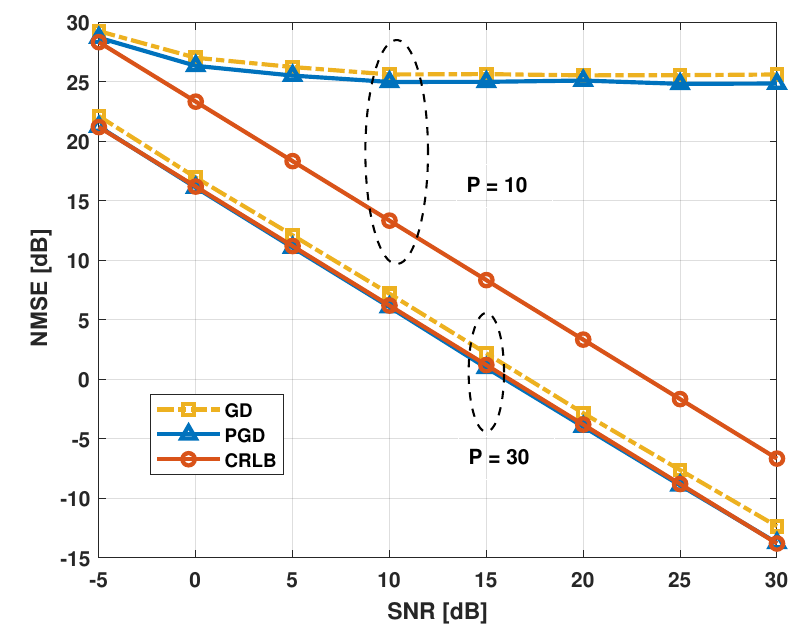}
	\caption{NMSE performance comparison with respect to the SNR under Rydberg atom-based 2D antenna array.}
    \label{NMSE_SNR_2D}
\end{figure}

However, the real benefit of GD method is not in 1D array, but in 2D array where GS algorithm is not applicable anymore. Fig. \ref{NMSE_SNR_2D} shows the NMSE performance for a Rydberg atom-based 2D antenna array, with $8 \times 8$ vapor elements.  When the pilot length is small ($P=10$), the PGD performs better than the basic GD, as the projection operation in PGD helps to exploit the rank constraint. However, their performance is far away from CRLB. On the other hand, when the pilot length is relatively large ($P=30$), both PGD and GD perform almost the same and close to the CRLB.

\begin{figure}[t]
	\centering
	\includegraphics[width=0.33\textwidth]{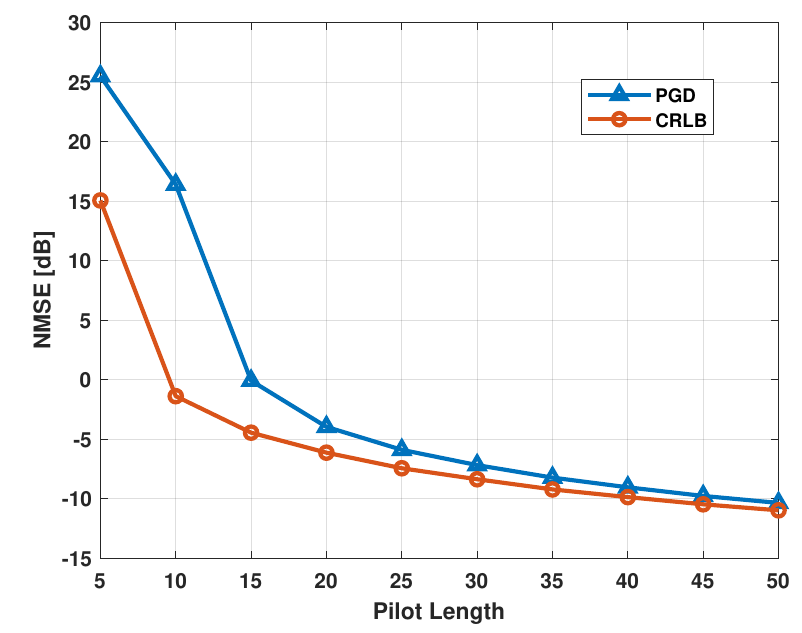}
	\caption{NMSE performance with respect to pilot length.}
    \label{NMSE_P}
\end{figure}

Fig. \ref{NMSE_P} depicts the NMSE comparison with respect to the pilot length $P$ when $\text{SNR}=5$ dB and in 2D array. 
It can be observed that the NMSE performance improves as the pilot length $P$ becomes longer.
In particular, to recover channel information with accuracy close to the CRLB, the length of the pilot sequence should be longer than 20, which is quite reasonable in practice.

\section{Conclusions}
In this letter, we proposed a gradient descent based method for channel estimation in Rydberg atomic receivers, and it can be effectively applied to both 1D and 2D antenna arrays. We reformulated the channel estimation problem using unfolded tensors, and CRLB has been derived to provide a performance benchmark for the estimation algorithm. It is found that when the number of pilots is limited, the proposed algorithm demonstrates a better performance than the phase retrieval algorithm in 1D case. Furthermore, when the pilot length reaches 20, the proposed algorithms can perform close to the CRLB in 2D antenna array.



\bibliographystyle{IEEEtran}
\bibliography{IEEEabrv,Ref}
\end{document}